\documentclass[12pt]{article}
\usepackage{amsfonts}
\usepackage{amsmath, graphicx, epsfig,psfrag, verbatim}
\usepackage{eucal,  amssymb,amsmath,amsthm,graphicx, amsfonts, latexsym}
\usepackage[utf8]{inputenc}

\begin{document} \parskip=5pt plus1pt minus1pt \parindent=0pt
\title{The disease-induced herd immunity level for Covid-19 is substantially lower than the classical herd immunity level}
\author{Tom Britton$^{1}$, Frank Ball$^{2}$ and Pieter Trapman$^{3}$}
\date{\today}
\maketitle

\begin{abstract}
Most countries are suffering severely from the ongoing covid-19 pandemic despite various levels of preventive measures. A common question is if and when a country or region will reach herd immunity $h$. The classical herd immunity level $h_C$ is defined as  $h_C=1-1/R_0$, where $R_0$ is the basic reproduction number, for covid-19 estimated to lie somewhere in the range 2.2-3.5 depending on country and region. It is shown here that the disease-induced herd immunity level $h_D$, after an outbreak has taken place in a country/region with a set of preventive measures put in place, is actually substantially smaller than $h_C$. As an illustration we show that if $R_0=2.5$ in an age-structured community with mixing rates fitted to social activity studies, and also categorizing individuals into three categories: low active, average active and high active, and where preventive measures affect all mixing rates proportionally, then the disease-induced herd immunity level is $h_D=43\%$ rather than $h_C=1-1/2.5=60\%$. Consequently, a lower fraction infected is required for herd immunity to appear. The underlying reason is that when immunity is induced by disease spreading, the
proportion infected in groups with high contact rates is greater than that in groups with low contact rates. Consequently, disease-induced immunity is stronger than when immunity is uniformly distributed in the community as in the classical herd immunity level.
\end{abstract}

\footnotetext[1]{Stockholm University, Department of Mathematics, Sweden. E-mail: tom.britton@math.su.se}
\footnotetext[2]{University of Nottingham, School of Mathematical Sciences, UK.}
\footnotetext[3]{Stockholm University, Department of Mathematics, Sweden.}

\section*{Introduction}\label{sec-Intro}

Covid-19 is spreading in most countries of the world and many different preventive measures are put in place to reduce transmission. Some countries aim for suppression by means of a total lockdown, and others for mitigation by slowing the spread using certain preventive measures in combination with protection of the vulnerable \cite{FMG20}. An important question for both policies is when to lift some or all of the restrictions. A highly related question is if and when herd immunity is obtained. 
Some regions and countries have already reached high estimates for the population immunity level, with 26\% infected in metropolitan Stockholm region as per May 1 2020 \cite{FHM20}, while by the end of March \cite{FMG20} estimates for Italy and Spain as a whole were already around or above 10\% and more recent. It is debated if the (classical) herd immunity level  $h_C=1-1/R_0$, which for Covid-19 is believed to lie between 50\% and 75\% since common estimates of $R_0$ for Covid-19 typically lie in the range 2-4 (e.g.\  \cite{FMG20}), is at all realistic to achieve without too many case fatalities \cite{FLN20,BAB20}

Herd immunity is defined as a level of population immunity such that disease spreading will decline and stop also after all preventive measures have been relaxed. If all preventive measures are relaxed when the immunity level (from people having been infected) is clearly below the herd immunity level, then a second wave of infection will start once restrictions are relaxed.

The classical herd immunity level $h_C$ is defined as $h_C=1-1/R_0$, where $R_0$ is the basic reproduction number defined as the average number of new infections caused by a typical infected individual during the early stage of an outbreak \cite{DHB13}. This definition originates from vaccination considerations: if a fraction $v$ is vaccinated (with a vaccine giving 100\% immunity) and vaccinees are selected uniformly in the community, then the new reproduction number is $R_v=(1-v)R_0$. From this it is clear that the critical vaccination coverage $v_c=1-1/R_0$; if at least this fraction is vaccinated, the community has reached herd immunity, as $R_v\le 1$, and no outbreak can take place.

\section*{An epidemic model for an age and activity-level structured population}

The simplest of all epidemic models is to assume a homogeneously mixing population in which all individuals are equally susceptible, and equally infectious if they become infected. We let $\lambda $ denote the average number of infectious contacts an individual who becomes infected has before recovering and becoming immune (or dying). An infectious contact is defined as one close enough to infect the other individual if this individual is susceptible (contacts with already infected individuals have no effect). For this simple model the basic reproduction number $R_0$ equals $R_0=\lambda$ \cite{DHB13}.

To this simple model we add two important features known to play an important role in disease spreading (the model is described in full detail in the Supplementary Information, SI). The first is to include age structure by dividing the community into different age cohorts, with inhomogeneous mixing between the different age cohorts. More precisely, we divide the community into 6 age groups and fit contact rates from an empirical study of social contacts \cite{WTK06}. Consequently, the person-to-person infectious contact rate between two individuals depends on the age groups of both individuals. The average number of infectious contacts an infected in age group $i$ has with individuals in age group $j$ now equals $\lambda a_{ij}\pi_j$, where $a_{ij}$ reflects both how much an $i$-individual has contact with a specific $j$-individual but also typical infectivity of $i$-individuals and susceptibility of $j$-individuals, and $\pi_j$ denotes the population fraction of individuals belonging to age cohort $j$.

The second population structure added categorizes individuals according to their social activity levels. A common way to do this is by means of network models (e.g.\ \cite{PV01}). Here we take a simpler approach and assume that individuals can be categorized into three different activity levels: 50\% of each age-cohort have normal activity, 25\% have low activity corresponding to half as many contacts compared to normal activity, and 25\% have high activity corresponding to twice as many contacts as normal activity. By this we mean that, for example, a high-activity individual in age group $i$ on average has $2*a_{ij}\pi_j*0.5*0.25$ infectious contacts with low-activity individuals of type $j$. The factor 2 comes from the infective having high activity, the factor 0.5 from the contacted person having low activity, and the factor 0.25 from low-activity individuals making up 25\% of each age cohort. The basic reproduction number $R_0$ for this model is given by the dominant eigenvalue to the (next-generation) matrix $M$ having these elements as its entries. \cite{DHB13}.

The final fraction getting infected in the epidemic, starting with few initial infectives, is given as the unique non-negative solution to a set of equations (the final-size equations) given in the SI. This solution also agrees with the final fraction infected of different types for the  corresponding stochastic epidemic model assuming a large population \cite{BC93}. In order to also say something about the time evolution of the epidemic we assume a classical SEIR epidemic model. More precisely, we assume that individuals who get infected are initially latent for a period with mean 3 days, followed by an infectious period having mean 4 days, thus approximately mimicking the situation for Covid-19 (e.g.\ \cite{FMG20}). During the infectious period an individual makes infectious contacts at suitable rates such that the mean number of infectious contacts agree with that of the next-generation matrix $M$.

\section*{The epidemic model with preventive measures put in place}

We assume that the basic reproduction number satisfies $R_0=2.5$ (a few other values are also evaluated) and that the epidemic is initiated with a small fraction of infectious individuals on February 15. On March 15, when the fraction infected is still small,  preventive measures are implemented such that all averages in the
next-generation matrix are scaled by the same factor $\alpha <1$, so the next-generation matrix becomes $\alpha M$. Consequently, the new reproduction number is $\alpha R_0$. These preventive measures are kept until the ongoing epidemic is nearly finished. More precisely, all preventive measures are relaxed thus setting $\alpha$ back to 1 on June 30. If herd immunity is not reached there will then be a second wave, whereas the epidemic continues to drop if herd immunity has been reached.

In the results section we investigate the effect of the preventive measures and for a couple of different scenarios numerically analyse whether or not a given level of preventive measures will yield disease-induced herd immunity. We do this for populations that a) are homogeneous population b) have individuals categorized into different age groups but no activity levels, c) have no age groups but different activity levels, and d)  have both age and activity structures.

\section*{Results}

For each of the four population structures,
we first show the overall disease-induced herd immunity level in Table \ref{tab_herd_levels}.
The level is obtained by assuming that preventive measures having factor $\alpha<1$ are implemented at the start of an epidemic, running the resulting epidemic to its conclusion and then exposing the population to a second epidemic with $\alpha=1$.  We find $\alpha_*$, the greatest value of $\alpha$ such that the second epidemic is subcritical; $h_D$ is then given by the fraction of the population that is infected by the first epidemic.  This approximated the situation where preventive measures are implemented early and lifted late in an outbreak. Note that $h_D$ is independent of the distributions of the latent and infectious periods.
\begin{table}[ht]
\caption{Disease-induced herd immunity level $h_D$ and classical herd immunity level $h_C=1-1/R_0$ for different population structures, for $R_0=2.0$, 2.5 and 3.0. Numbers correspond to percentages. }
\centering 
\begin{tabular}{| c || c | c || c | c || c|c |}
\hline             
Population structure & $h_D$ & $h_C$ & $h_D$ & $h_C$ & $h_D$ & $h_C$
\\
[0.5ex]
\hline                  
Homogeneous & 50.0 & 50.0  & 60.0 & 60.0 & 66.7 & 66.7
\\
Age structure & 46.0 & 50.0 & 55.8 & 60.0 & 62.5 & 66.7
\\
Activity structure & 37.7 & 50.0   & 46.3 & 60.0 & 52.5 & 66.7
\\
Age \& Activity structure & 34.6 & 50.0  & 43.0 & 60.0 & 49.1 & 66.7
\\
[1ex]      
\hline
\end{tabular}\label{tab_herd_levels}
\end{table}
\bigskip

As seen in the table all three structured population have lower disease-induced herd immunity $h_D$  compared to the classical herd immunity $h_C$, which assumes immunity is uniformly distributed among the different types of individual. From the table it is clear that the different activity levels have a greater effect on reducing $h_D$ than age structure.

In Table \ref{tab_fin_sizes} the final fractions infected in the different age activity groups for $\alpha=\alpha_*$ just barely reaching disease-induced herd immunity are given. This is done for the age and activity group structure and assuming $R_0=2.5$. The overall fraction infected equals $h_D=43.0\%$, in agreement with Table \ref{tab_herd_levels}.

\begin{table}[ht]
\caption{Final outcome fractions infected in different groups assuming $R_0=2.5$ and preventive measures put in place such that $\alpha=\alpha_*$ just barely reaching herd immunity for  $R_0=2.5$. Population structure includes both age and activity and fractions infected are given as percentages.}
\centering 
\begin{tabular}{| c | c | c | c | }
\hline             
Age-group & Low activity & Average activity & High activity
\\
[0.5ex]
\hline                  
0 - 5 years &  17.6   & 32.1  &  53.9
\\
6 - 12 years &     25.8  &  44.9  &  69.7
\\
13 - 19 years &     31.4   & 52.9   & 77.8
    \\
20 - 39 years &     27.4  &  47.2  &  72.1
    \\
40 - 59 years &     22.8  &  40.3  &  64.4
    \\
$\ge$ 60 years &     14.6  &  27.0  &  46.7
\\
[1ex]      
\hline
\end{tabular}\label{tab_fin_sizes}
\end{table}
\bigskip

\begin{figure}[ht!]
    \centering
  	\includegraphics[width=1\textwidth, height=1\textheight, angle=0]{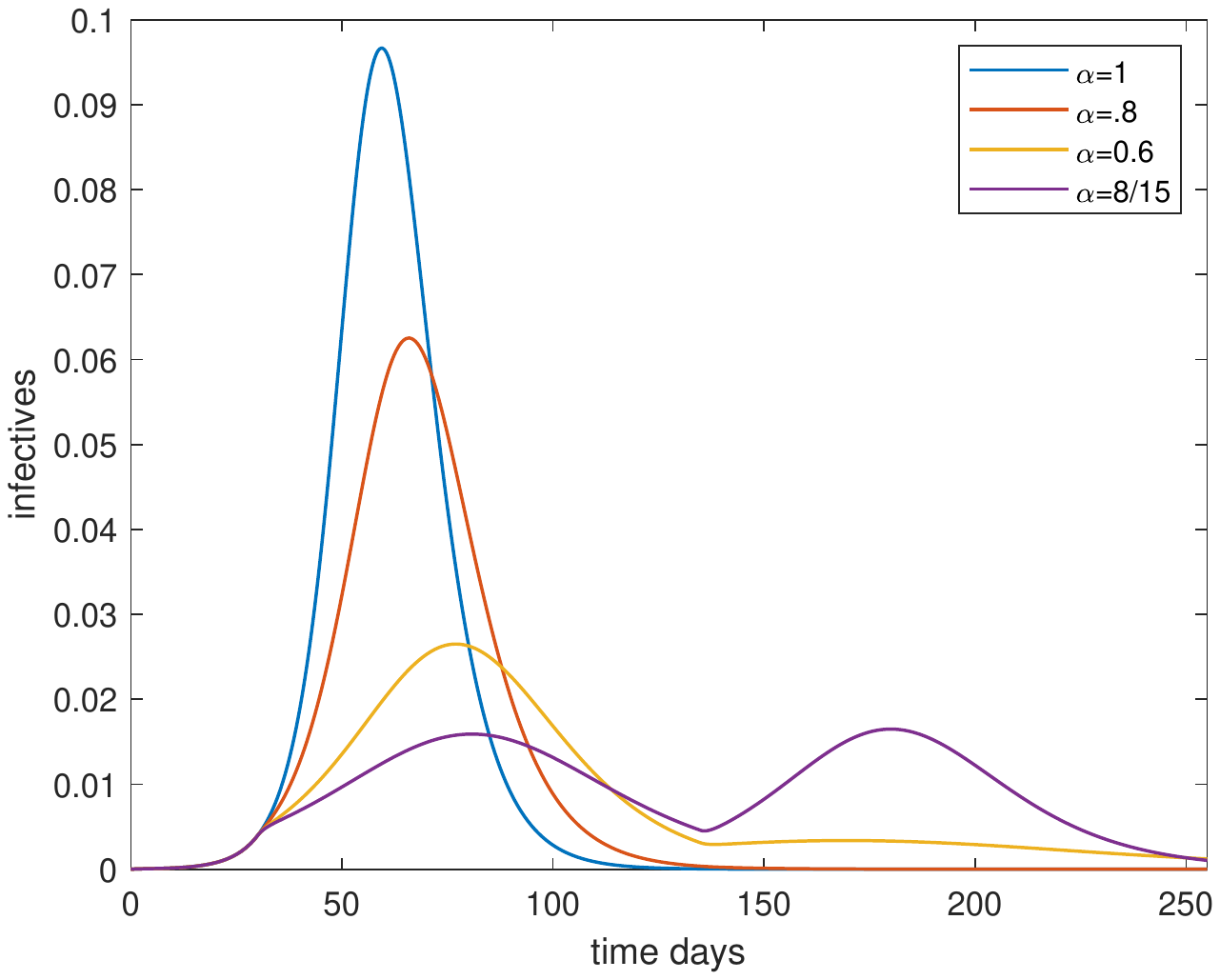}
  	\vskip-7cm
   \caption{Plot of the overall fraction infected over time for the age and activity structured community with $R_0=2.5$, for four different preventive levels inserted March 15 (day 30) and lifted June 30 (day 135). The black, red, yellow and purple curves corresponds to no, light, moderate and severe preventive measures, respectively.}
\label{Fig_incidence}
\end{figure}

We also illustrate the time evolution of the epidemic for $R_0=2.5$, assuming both age and activity structure, and starting with a small fraction externally infected in mid-February. For this we show the epidemic over time for four different levels of preventive measures put in place early in the epidemic outbreak (mid-March) and being relaxed once transmission has dropped to low levels (June 30). In Figure \ref{Fig_incidence} the community proportion that is infectious is plotted during the course of the epidemic.

It is seen that preventive measures reduce the size and delay the time of the peak. On March 15 preventive measures (at four different levels for $\alpha$) are put in place and it is seen that the growth rate is reduced except for the black curve which has no preventive measures ($\alpha=1$). Sanctions are lifted on June 30 putting transmission rates back to their original levels, but only in the curve with highest sanctions is there a clear second outbreak wave, since the remaining curves have reached (close to) herd immunity. The yellow curve finishes below 50\% getting infected. The reason it has more than the 43\% infected shown in Table \ref{tab_herd_levels} is that preventive measures were not from the very start and were also lifted before the epidemic was over. An interesting observation is that the purple curve ends up with a \emph{higher} overall fraction infected even though it had \emph{more restrictions} than those of the yellow. The explanation is that this epidemic was further from completion when sanctions were lifted.

\begin{figure}[ht!]
    \centering
  	\includegraphics[width=1\textwidth, height=1\textheight, angle=0]{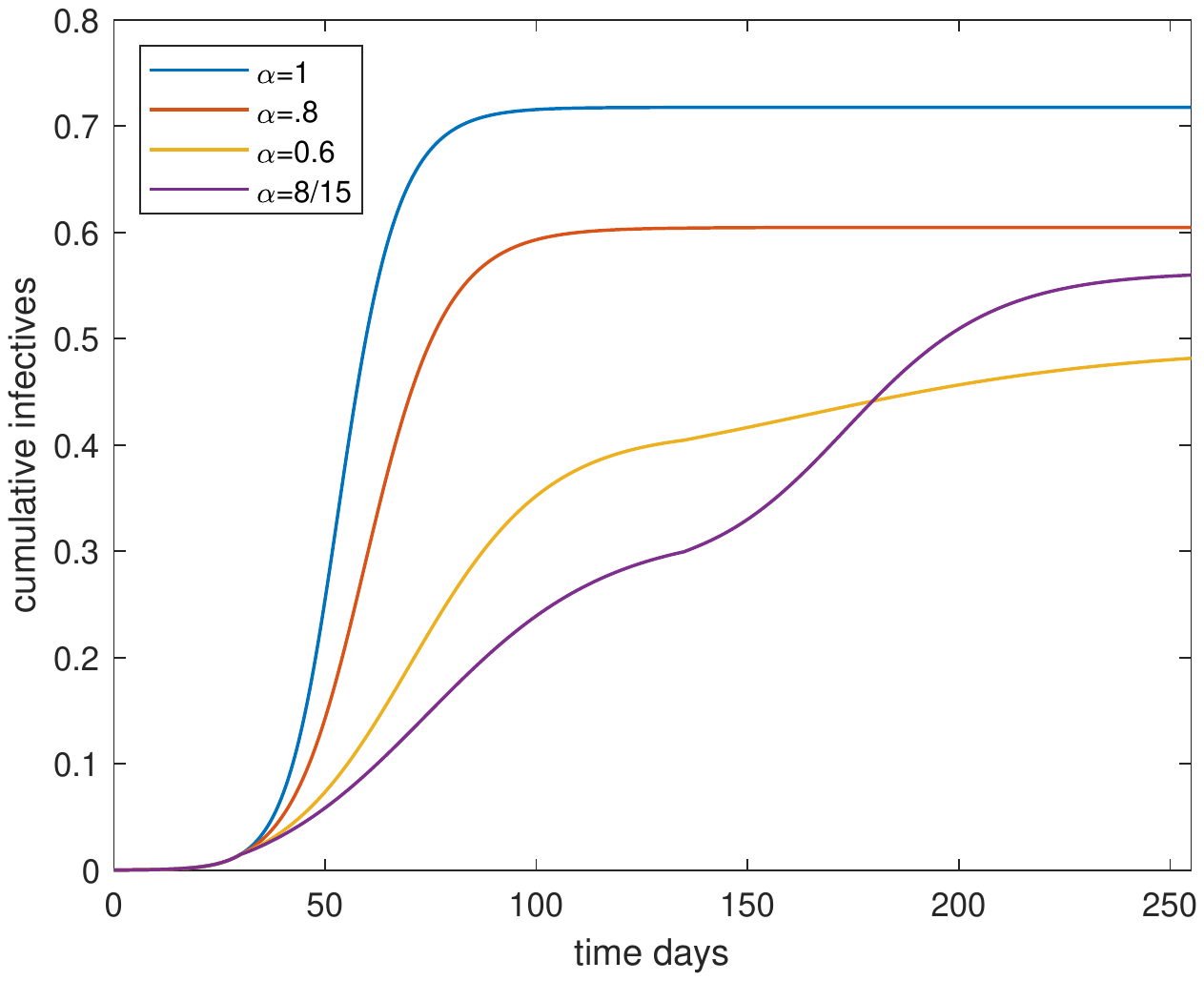}
 	\vskip-7cm
   \caption{Plot of the cumulative fraction infected over time for the age and activity structured community and $R_0=2.5$, for a four different preventive levels inserted March 15 and lifted June 30. The black curve corresponds to no preventive measures, the red with light preventive measure, the yellow to moderate preventive measures and the purple corresponding to severe preventive measures.}\label{Fig_cumulative}
\end{figure}

Figure \ref{Fig_cumulative} plots the corresponding cumulative fraction infected as a function of time.

Observe that the first three curves see no (strong) second wave of outbreak once preventive measures are lifted -- it is only the curve corresponding to highest preventive measures that has a severe second wave.
when restrictions are lifted.
Note that also the yellow curve, having overall fraction infected well below the classical herd immunity level $h_C=60\%$ is protected by herd immunity since no second wave appears. This clearly illustrates that the disease-induced herd immunity level $h_C$ is well below 60\% -- it is 43\% (see Table~\ref{tab_herd_levels}). In the SI we show additional plots for the situation where restrictions are lifted continuously between June 1 and August 31, and also study how the effective reproduction number evolves as a function of the time when restrictions are lifted.

\section*{Discussion}

The main conclusion is that the disease-induced herd immunity level may be substantially lower then the classical herd immunity level. Our illustration indicates a reduction from 60\% down to 43\% (assuming $R_0=2.5$) but this should be interpreted as an illustration, rather than an exact value or even a best estimate.

The current model took age cohorts and social activity levels into account. However, more complex and realistic models have many other types of heterogeneities: for instance increased spreading within households (of different sizes) or within schools and workplaces; and spatial aspects with rural areas having lower contact rates than metropolitan regions. It seems reasonable to assume that most such additional heterogeneities will have the effect of \emph{lowering} the disease-induced immunity level $h_DS$ even further, in that high spreading environments (metropolitan regions, large households, big workplaces, ...) will have a higher fraction infected thus resulting in immunity being concentrated even more on highly active individuals. Some complex models do not categorize individuals into different activity levels, or the related feature in an underlying social network with varying number of acquaintances. As illustrated in our results section, differences in social activity plays a greater role in reducing the disease-induced herd immunity level than inhomogeneous age-group mixing. Thus models not having such features will see smaller difference between $h_D$ and $h_C$. Our choice to have 50\% having average activity, 25\% having half and 25\% having double activity is of course very arbitrary. An important future task is hence to determine how size of differences in social activity within age groups. The more social heterogeneity there is between groups, the bigger difference between $h_D$ and $h_C$.

An assumption of our model is that preventive measures acted proportionally on all contact rates. This may not always hold. For example, most countries have a pronounced ambition to protect elderly (and other risk groups), which does not obey this assumption. Again, we expect the effect would be to \emph{lower} the disease-induced immunity level had this type of preventive measure been considered, because the oldest age group is the one having fewest contacts. For a model having schools and workplaces, it is however not obvious what effect school closure and strong recommendations to work from home would have on the disease-induced herd immunity level.

There are of course other more efficient exit-strategies than to lift all restrictions simultaneously. In fact, most countries are currently employing a gradual lifting of preventive measures. Such slower lifting of preventive measures will avoid seeing the type of overshoot illustrated by the purple curve in Figure \ref{Fig_cumulative}, which results in a greater fraction infected than the yellow curve, even though the latter has milder restrictions. The effect of such gradual lifting of restrictions will result in the final fraction infected reaching close to the disease-induced herd immunity level.

Different forms of immunity levels have been discussed previously in the literature although, as far as we know, not when considering early-introduced preventions that are lifted towards the end of an epidemic outbreak. Anderson and May \cite{AM91} concludes that immunity level may differ between uniformly distributed, disease-induced and optimally-located immunity, and vaccination policies selecting individuals to immunize in an optimal manners have been discussed in many papers, e.g.\ \cite{BBL04}. A very recent and independent preprint make similar observations to those in the present paper, but where heterogeneities in terms of e.g.\ susceptibilities vary continuously \cite{GAC20}. The correspondning epidemic model is solved numerically similar to our Figure \ref{Fig_incidence}, but analytical results for final sizes and $h_C$ are missing.

The present study highlights that the disease-induced herd immunity level $h_D$ is substantially smaller than the classical herd immunity level $h_C$. To try to quantify more precisely the size of this effect remains to be done, and we encourage more work in this area.


\bibliography{herdref}
\bibliographystyle{ieeetr}

\pagebreak

\centerline{\textbf{\Large Supplementary information}}


\section*{Materials and Methods}

\subsection*{A deterministic SEIR model and the fraction of the population infected}

In this supplementary information we describe the deterministic SEIR (Susceptible, Exposed, Infectious, Removed) epidemic model in a population partitioned by age and activity level. For reasons of notational convenience we label the types (the  combination of age and activity level) from $1$ to $m$, where $m$ is the product of the number of age classes and the number of activity levels. A more detailed exposition as the one presented here can be found in [1, Sections 5.5 and 6.2].

We assume that for all $j \in \{1,\cdots,m\}$ the population consists of $n_j$ people of type $j$.  We set $n = \sum_{j=1}^m n_j$ and $\pi_j = n_j/n$. We assume that the population is large and  closed, in the sense that we do not consider births, deaths (other than possibly the deaths caused by the infectious disease) and migration. Throughout the epidemic $n_i$ is fixed. So, people who die by the infectious disease are still considered part of the population.  For $j,k \in \{1,\cdots,m\}$, every given person of type $j$ makes infectious contacts with every given person of type $k$ independently at rate $\alpha a_{jk}/n$. If at the time of such a contact the type $j$ person is infectious and the type $k$ person is susceptibe then the latter becomes latently infected (Exposed). People of the same type may infect each other, so $a_{jj}$ may be strictly positive. Because the definition of an infectious contact includes that the contact leads to transmission of the disease, it is not necessarily the case that $a_{jk}$ is equal to $a_{kj}$. The parameter $\alpha$ is a scaling parameter, used to quantify the impact of control measures in the main paper, without measures $\alpha$ is set equal to 1. Exposed individuals become Infectious at constant rate $\sigma$ and infectious individuals recover or die (are Removed) at constant rate $\mu$. The rates of becoming infectious and removal are assumed to be independent of type. It is straightforward to extend the model to make those rates  age or activity level dependent. 

In the described multi-type SEIR model, the expected number of people of type $k$  that are infected by an infected person of type $j$ during the early stages of the infection is $n_k \times (\alpha a_{jk}/n) \times (1/\mu) = \pi_k \alpha a_{jk}/\mu$, where $1/\mu$ is the expected duration of an infectious period. The next generation matrix $M$ has for $j,k \in \{1,\cdots,m\}$ as element in the $j$-th row and $m$-th column the quantity  $\pi_k \alpha a_{jk}/\mu$. We define the basic reproduction number, $R_0$ as the largest eigenvalue of $M$, which is necessarily real and positive. If $R_0>1$, then a large outbreak is possible with positive probability, while if $R_0 \leq 1$ an outbreak stays small with probability 1.

We set $S_j(t)$ to be the number of people of type $j$ that are susceptible to the disease at time $t$, $E_j(t)$ the number of people of type $j$ that are latently infected, $I_j(t)$ the number of infectious people of type $j$ and $R_j(t)$ the number of removed people of type $j$ ($j \in \{1,\cdots,m\}$). Note that $S_j(t)+E_j(t)+I_j(t)+R_j(t) = n_j = \pi_j n$ for all $t\geq 0$, because the population is closed. 
Again for $j \in \{1,\cdots,m\}$ we define $s_j(t)=S_j(t)/n_j$, $e_j(t)=E_j(t)/n_j$, $i_j(t)=I_j(t)/n_j$ and $r_j(t)=R_j(t)/n_j$. 

Theory on Markov processes  [2, Chapter 11] (see also  [1, Section 5.5] for the single type counterpart) gives that for large $n$ the above model can be described well by a system of differential equations (again for $j \in \{1,\cdots,m\}$).
\begin{displaymath}
\begin{array}{rllll}
\dot{s}_j(t) & = & - \frac{1}{n_j} \displaystyle\sum_{k=1}^m \alpha \frac{a_{kj}}{n} S_j(t) I_k(t) &= &
-\displaystyle\sum_{k=1}^m \alpha \pi_k a_{kj} s_j(t) i_k(t),\\
\dot{e}_j(t) & = & \frac{1}{n_j}\left( \displaystyle\sum_{k=1}^m \alpha \frac{a_{kj}}{n} S_j(t) I_k(t) - \sigma  E_j(t) \right)&= & 
\displaystyle\sum_{k=1}^m \lambda \pi_k a_{kj} s_j(t) i_k(t) -\sigma e_j(t),\\
\dot{i}_j(t) & = &  \frac{1}{n_j}\left( \sigma  E_j(t) - \mu  I_j(t) \right) &= & 
\sigma e_j(t)-\mu i_j(t),\\
\dot{r}_j(t) & = &  \frac{1}{n_j} \mu I_j(t)  &= & \mu i_j(t).
\end{array}
\end{displaymath}
To be complete we use for the analysis in the main text that for some $j_* \in \{1,\cdots,m\}$, $s_{j_*}(0)=1-\epsilon$, $e_{j_*}(0)=\epsilon$ and $i_{j_*}(0)=r_{j_*}(0)=0$ while  $s_{j}(0)=1$, $e_{j}(0)=i_{j}(0)=r_{j}(0)=0$  for $j \in \{1,\cdots,m\}\setminus j_*$. In the analysis below we do not impose assumptions on the initial conditions.

Because the population is closed, the epidemic will ultimately go extinct and therefore for all $j \in \{1,\cdots,m\}$ we have that  $e_j(t) \to 0$ and $i_j(t) \to 0$ as  $t \to  \infty$. So $s_j(t) + r_j(t) \to 1$  as $t \to \infty$. Furthermore $s_j(t)$ is non-increasing. Therefore $s_j(\infty) = \lim_{t \to \infty} s_j(t)$ exists.

It can be shown in the spirit of [1, Equation (6.2)] that for $j \in \{1,\cdots,m\}$, 
\begin{equation}
\label{finalsize}
\frac{s_j(\infty)}{s_j(0)} = \exp\left[-\lambda \sum_{k=1}^m a_{kj} \pi_k \left(1-r_k(0)-s_k(\infty)\right)/\mu\right].
\end{equation} 
To understand this identity  we observe first that $\frac{s_j(\infty)}{s_j(0)}$ is the fraction of initially susceptible people of type $j$ escapes the epidemic.
while the exponent in the right hand side can be written as
$$\sum_{k=1}^m n \pi_k \left(1-r_k(0)-s_k(\infty)\right) \times \lambda a_{kj}/n \times \frac{1}{\mu}
=  \sum_{k=1}^m (n_k-R_k(0)-S_k(\infty)) \times \lambda a_{kj}/n \times \frac{1}{\mu}.
$$ 
In words the summands read as the number of people of type $k$ that were infectious at some moment during the epidemic, times the rate at which a type $k$ person makes infectious contacts with someone of type $j$,  times the expected time an infected person is infectious. In other words, the right hand side is the cumulative force of infection during the entire epidemic on a person of type $j$. Standard theory on epidemics gives that minus the natural logarithm of the probability that a given initially susceptible person of type $j$ avoids infection is the cumulative force of infection on the person. 

So \eqref{finalsize} gives that the fraction of initially susceptible people that are ultimately still susceptible is equal to the probability that a given initially susceptible person avoids infection. 

\subsection*{The population matrix}

In the main text we analyse an age structured population. Contact intensities between different age groups we took from [3]. The age groups are 0-5, 6-12, 13-19, 20-39, 40-59 and 60+.   The contact matrix, i.e.\ the matrix with elements $\{a_{jk};j,k \in\{1,\cdots,6\}\}$ is taken from Table 1 of [3]. Note that the expected number of contacts from a person of type $j$ with people of type $k$ is $n_k a_{jk}/n = \pi_k a_{jk}$. Therefore we  divide the elements of Table 1 by $\pi_k$ to obtain the contact matrix. We further multiply this matrix by a constant such that the largest eigenvalue is equal to $2.5$,t he value we have chosen for $R_0$.
The contact matrix is
$$
\begin{pmatrix}
 2.2257  &  0.4136 &   0.2342  &  0.4539   & 0.2085  &  0.1506\\
    0.4139 &   3.6140 &   0.4251 &   0.4587  &  0.2712  &  0.1514\\
    0.2342 &   0.4257 &   2.9514  & 0.6682 &    0.4936  &  0.1972\\
    0.4539  &  0.4592  &  0.6676 &   0.9958  &  0.6510   & 0.3300\\
    0.2088 &   0.2706  &  0.4942  &  0.6508 &  0.8066  &  0.4341\\
    0.1507 &   0.1520   & 0.1968 &   0.3303 &   0.4344  &  0.7136
\end{pmatrix}.
$$
As explained in the main text we can use this matrix to generate the 18 by 18 contact matrix for the model in which we take both age and activity level into account. 

 \subsection*{Additional figures}
 
\begin{figure}[ht!]
    \centering
  	\includegraphics[width=1\textwidth, height=1\textheight, angle=0]{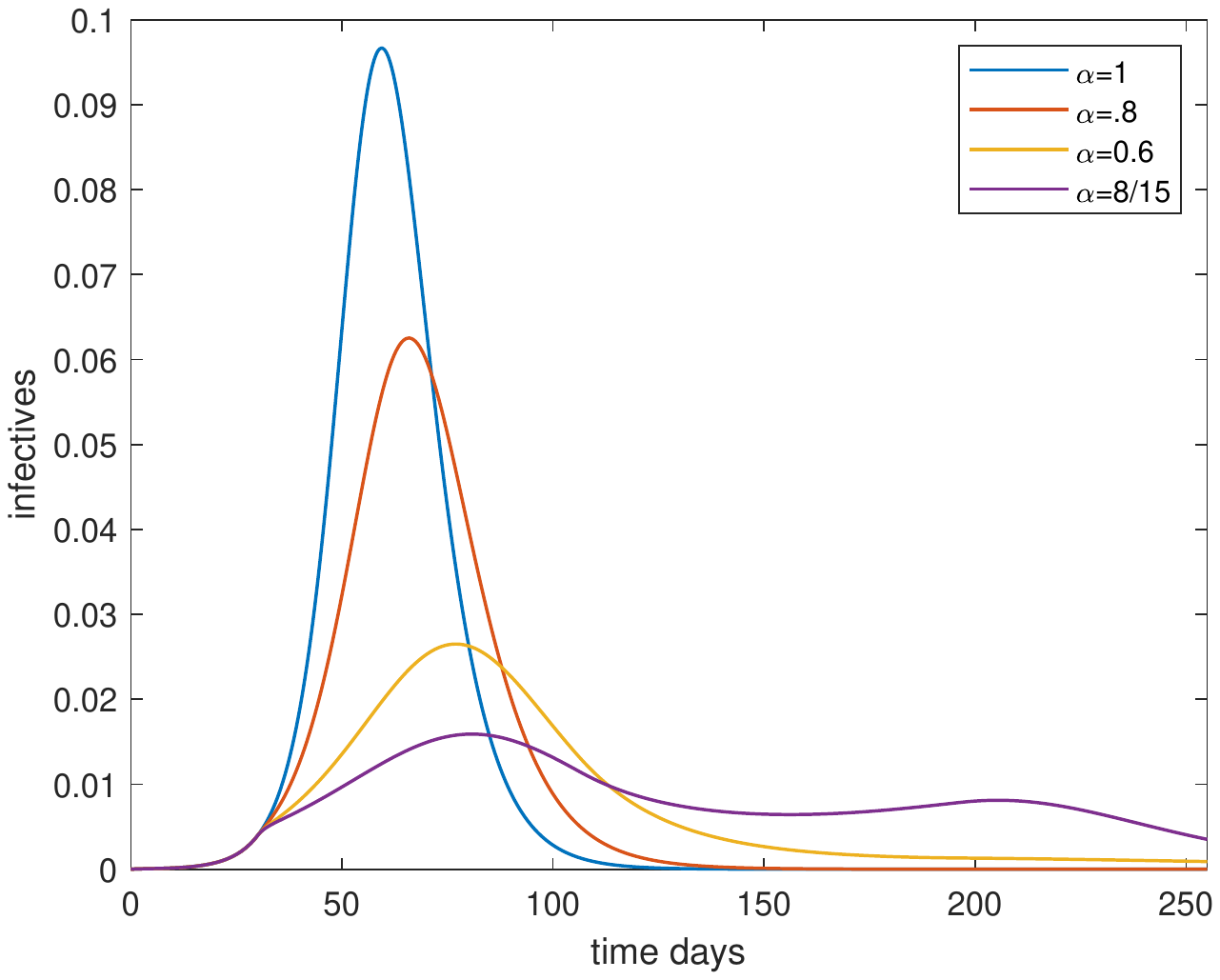}
  	\vskip-7cm
   \caption{Plot of the overall fraction infected over time for the age and activity structured community with $R_0=2.5$, for four different preventive levels inserted March 15 (day 30) and lifted gradually between June 1 (day 105) and August 31 (day 195). The black, red, yellow and purple curves corresponds to no, light, moderate and severe preventive measures, respectively.}
\label{Fig_Gradual_inc}
\end{figure}

In the main text we studied effects of lifting restrictions of different levels $\alpha$ on June 30 (day 135) going back to the situation of no restrictions corresponding to setting $\alpha$ back to 1. Below are the corresponding plots but where restrictions are relaxed gradually (linearly) between  June 1 (day 105) and August 31 (day 195). In Figure \ref{Fig_Gradual_inc} the we plot the fraction of infectious individuals.

The plot looks quite similar to that of Figure \ref{Fig_incidence} in the main text except that the purple curve with highest restrictions no longer has a pronounced second wave. The reason for this is that now restrictions are lifted slowly and graduella during a 3 month period rather than discretely back to normal from one day to another.
\begin{figure}[ht!]
    \centering
  	\includegraphics[width=1\textwidth, height=1\textheight, angle=0]{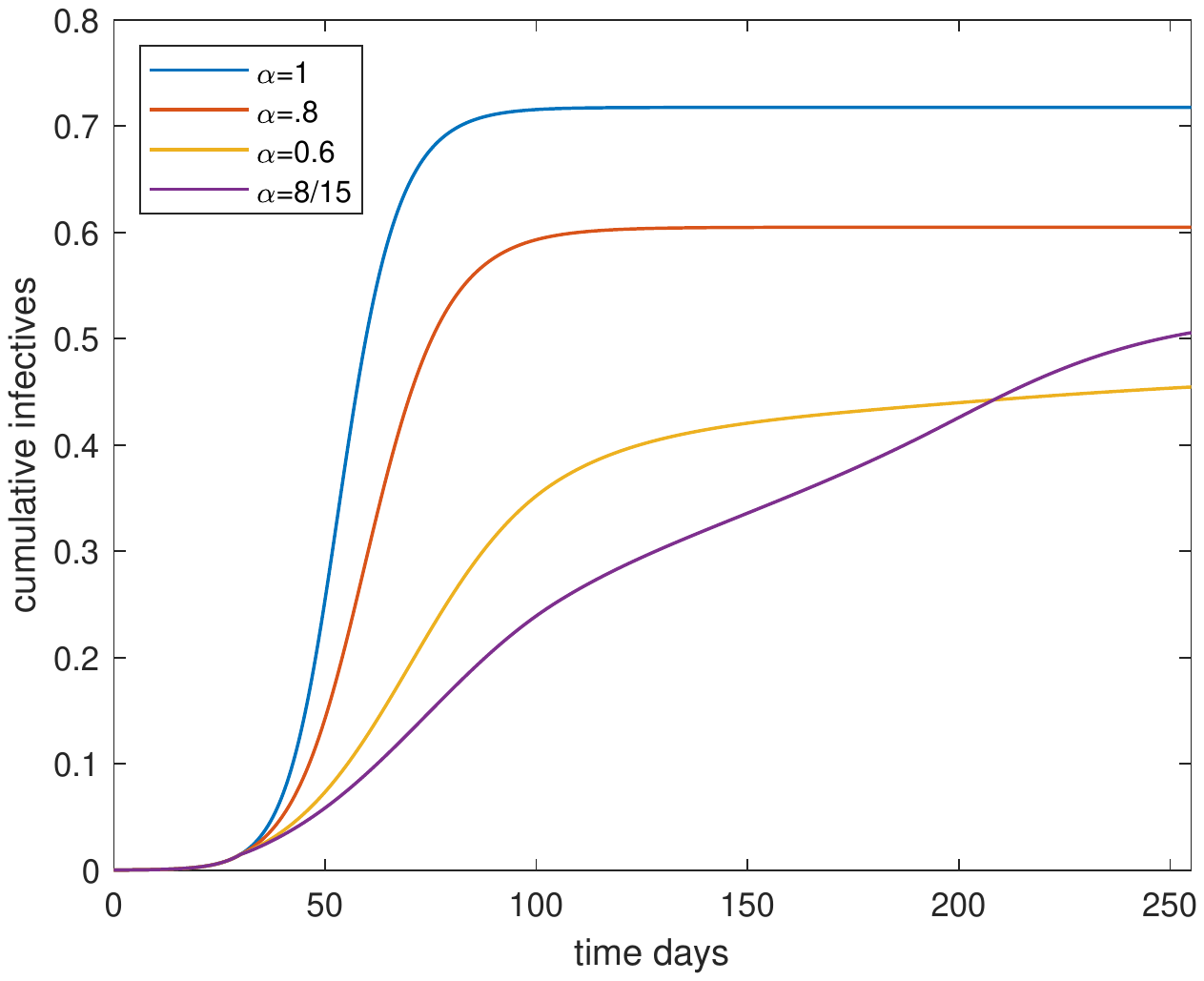}
  	\vskip-7cm
   \caption{Plot of the overall fraction infected over time for the age and activity structured community with $R_0=2.5$, for four different preventive levels inserted March 15 (day 30) and lifted gradually between June 1 (day 105) and August 31 (day 195). The black, red, yellow and purple curves corresponds to no, light, moderate and severe preventive measures, respectively.}
\label{Fig_Gradual_cdf}
\end{figure}
In Figure \ref{Fig_Gradual_cdf} the corresponding plot for cumulative fraction of infected over time is given. Compared to Figure \ref{Fig_cumulative} tha main difference is that the purple curve (severe restrictions) has fewer finally infected since there is no longer as big over-shoot above $h_C=43\%$ caused by a second wave.

\begin{figure}[ht!]
    \centering
  	\includegraphics[width=1\textwidth, height=1\textheight, angle=0]{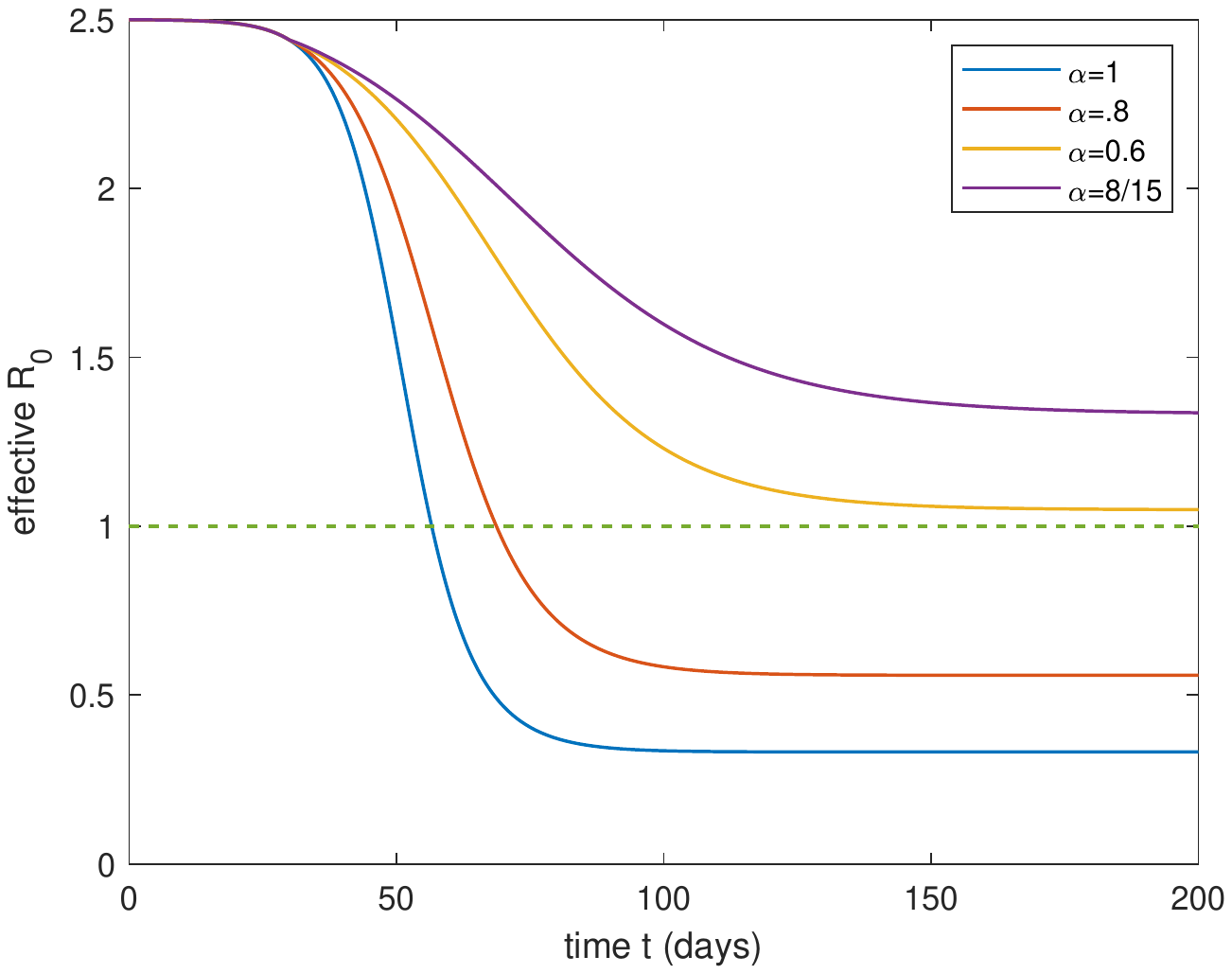}
  	\vskip-7cm
   \caption{Plot of the effective reproduction number (incorporating disease-induced immunty) if restrictions for different $\alpha$ are put in place Day 30 and relaxed on day $t>30$.}
\label{Fig_Eff_R}
\end{figure}

Finally, in Figure \ref{Fig_Eff_R} we consider the situation when preventive measures with level $\alpha$ are implemented 30 days after introduction of the disease and relaxed (so $\alpha$  returns to 1)
at time $t>30$.  The parameters are again chosen so that $R_0=2.5$ when $alpha=1$.  The graphs show the effective $R_0$ (incorporating disease-induced immunity) as a function of time $t$, for four different choices of preventive level $alpha$.
Thus all four curves coincide until day 30.  The effective $R_0$ with no preventive measures ($\alpha=1$)  reaches the critical value of one on about day 57 (mid-April), whilst that for $\alpha=0.8$ does so on about day $68$ (April 23).  The stronger preventive measures ($\alpha=0.6$ and $\alpha=8/15$) are such that herd immunity is never reached even if they are retained indefinitely.

 \section*{Additional References}
 
[1] H. Andersson and T. Britton, \emph{Stochastic epidemic models and their statistical analysis}. New York: Springer Verlag, 2000.

[2] S.N. Ethier and T.G. Kurtz. \emph{Markov processes: characterization and convergence}, vol. 282. John Wiley \& Sons, 2009.

[3] J. Wallinga, P. Teunis and M. Kretzschmar. Using data on social contacts to estimate age-specific transmission parameters for respiratory-spread infectious agents. \emph{American Journal of Epidemiology}. Vol 164: 936-944, 2006.

\end{document}